\documentclass{appolb}
\usepackage{epsfig}
\usepackage{cite}
\usepackage{mcite}
\begin{document}
\title{$D^{*\pm}$ Production in $e^-p$ and $e^+p$ \\ 
                   Deep Inelastic Scattering at HERA%
\thanks{Presented at DIS02}%
}
\author{S. D. Robins
\address{on behalf of the ZEUS collaboration}
%\and
%the Name(s) of other Author(s)
%\address{and their affiliation}
}
\maketitle
\begin{abstract}
Inclusive production of $D^{*\pm} (2010)$ mesons in deep inelastic scattering has been measured using $e^+p$ and $e^-p$ data obtained with the ZEUS detector at HERA using integrated luminosities of $16.7$ and $65.2 \rm pb^{-1}$, respectively.
The decay channel $D^{*+} \rightarrow D^{0} \pi^{+}$ with $D^{0} \rightarrow K^{-} \pi^{+}$ and corresponding antiparticle decays were used to identify $D^{*\pm}$ mesons. 
The $D^{*\pm}$ cross sections in $e^-p$ and $e^+p$ interactions agree with NLO QCD predictions, although the $D^{*\pm}$ cross section in $e^-p$ is slightly higher than that in $e^+p$.

\end{abstract}
%\PACS{}

\section{Introduction}
Charm production in deep inelastic scattering (DIS) at HERA has been
shown in previous studies to be consistent with purely dynamic
Boson-Gluon Fusion (BGF) production
\cite{epj:c12:35,zfp:c72:593,pl:b407:402}.  This agreement has now
been tested with a larger data sample than the previous ZEUS
measurements, and $e^-p$ as well as $e^+p$ cross sections have been
calculated.  The charmed mesons were identified using the decay
$D^{*+} \rightarrow D^{0} \pi^{+}$ with $D^{0} \rightarrow K^{-}
\pi^{+}$ and corresponding antiparticle processes, where $\pi_s$
refers to a low momentum pion accompanying the $D^0$.  The
differential cross sections are measured as functions of $Q^2$ and
Bjorken $x$, defined as $Q^2 = -q^2 = (k-k^{\prime})^2$ and $x = Q^2 /
(2P \cdot q)$, where $k$ and $k^{\prime}$ are the four-momenta of the
initial and final state lepton, and P is the four-momentum of the
proton.

\section{Kinematic reconstruction and event selection}
The ZEUS detector is described in detail elsewhere
\cite{pl:b293:465,zfp:c63:391}.  The $x$ and $Q^2$ variables were
reconstructed by the $\Sigma$-method, which uses both the scattered
lepton and the hadronic system measurements \cite{nim:a361:197}.
Standard cuts were imposed to select neutral current DIS events
\cite{misc:eps01:493}.  The $D^{\ast \pm}$ mesons were selected in the
range $1.80 < M(D^0) < 1.92$ GeV, $0.143 < \Delta M < 0.148$ GeV, $1.5
< p_T(D^\ast) < 15$ GeV, and $|\eta(D^\ast)| < 1.5$.  The number of
$D^{\ast\pm}$ events determined from a 5 parameter fit

\begin{center}
  $F( \Delta M) = [P1/\sqrt{2 \pi} \cdot P3] \cdot exp[0.5 \cdot(
  \Delta M - P2)^2 / P3] + P4 \cdot (\Delta M - m_{\pi}) ^{P5}$
\end{center}

\noindent
where $P1$ - $P5$ are free parameters is $1229 \pm 48$ in the $e^-p$
data, and $4240 \pm 90$ in the $e^+p$ data.  The number of $D^{\ast
  \pm}$ mesons extracted from empirical wrong charge background
subtraction within the signal region $143 < \Delta M < 148$ MeV, is
$1219 \pm 58$ and $4239 \pm 113$ in the $e^-p$ and $e^+p$ data
respectively.  The $\Delta M$ distributions are shown in Figure 1, for
$e^-p$ and $e^+p$ data separately.

\section{Study of systematic effects}

The systematic uncertainties on the measured $D^{\ast \pm}$ cross
sections were determined by changing the selection cuts or analysis
procedure.  These uncertainties are divided into three groups.\\

\noindent
Event reconstruction and selection:

\begin{itemize}
\setlength{\itemsep}{-5pt}
\item{The cuts on $y_e$. $y_JB$, $\delta$, and the vertex position were varied \cite{misc:eps01:493}.}
\item{The cut on the position of the scattered lepton in the RCAL was raised.}
\item{The minimum energy of the scattered lepton was raised.}
\item{The Electron method or Double Angle method \cite{proc:hera:1991:23} was used to reconstruct the kinematics.}
\end{itemize}
\noindent
$D^{\ast \pm}$ reconstruction:
\begin{itemize}
\setlength{\itemsep}{-5pt}
\item{A higher track quality was required (restriction on the polar angle of the track).}
\item{The transverse momentum requirement of the $K$ and $\pi$ candidates was varied.}
\item{The signal region for $M(D^0)$ and $\Delta M$ were varied.}
\end{itemize}
\noindent
Monte Carlo:
\begin{itemize}
\setlength{\itemsep}{-5pt}
\item{The acceptance was calculated using HERWIG \cite{cpc:67:465} instead of RAPGAP \cite{cpc:86:147}.}
\end{itemize}

\noindent
The overall systematic uncertainty was determined by adding the above
uncertainties in quadrature.  The normalisation uncertainties due to
the luminosity measurement error, and those due to the $D^{\ast \pm}$
and $D^0$ branching ratios were not included.

\section{Results}

In the kinematic region $1 < Q^2 < 1000$ GeV$^2$, $0.02 < y < 0.08$
and the selected $D^{\ast\pm}$ region, the cross sections calculated
are

\begin{center}
  $ \sigma (e^-p \rightarrow e^- D^{\ast\pm} X) = 10.20 \pm 0.48
  (stat.) ^{0.36} _{0.54} (syst.)$ nb,
\end{center}
\begin{center}
  $ \sigma (e^+p \rightarrow e^+ D^{\ast\pm} X) = 8.94 \pm 0.24
  (stat.) ^{0.27} _{0.51} (syst.)$ nb.
\end{center}

\noindent
The $e^+p$ cross section is consistent with that previously published
\cite{epj:c12:35}, allowing for the increase in proton beam energy
\cite{misc:eps01:493}, while the $e^-p$ cross section is slightly
higher.  Figure 2 shows the differential cross sections as a function
of $Q^2$ and $x$ compared to the NLO calculation implemented in the
HVQDIS program \cite{pl:b353:535,pr:d57:2806}.  This program is based
on the BGF mechanism, and uses the Peterson fragmentation function
\cite{pr:d27:105}, with $ \epsilon $ = 0.035, to hadronise the charm
quark to a $D^{\ast\pm}$.  The mass and renormalisation scales were
set to $\sqrt{ 4 m_c^2 + Q^2}$.  The hadronisation fraction $f (c
\rightarrow D^{\ast+})$ was set to 0.235 \cite{hep-ex-9912064}.  The
boundaries of the shaded band indicate two extreme values of HVQDIS
predictions, from changing the charm mass between $m_c = 1.3$ to $1.6$
GeV, and using different sets of structure functions, GRV98HO
\cite{epj:c5:461}, CTEQ5F3 \cite{epj:c12:375} and a ZEUS NLO fit
\cite{epj:c7:609}.  The NLO calculations based on BGF give a good
description of the measured $D^{\ast \pm}$ cross section over the full
range of $Q^2$ and $x$.  For $Q^2 > 20$ GeV$^2$, the $D^{\ast \pm}$
cross sections in $e^-p$ and $e^+p$ differ slightly, while
conventional charm production mechanisms contain no charge dependence
on the lepton in these interactions.  More $e^-p$ data is essential to
investigate whether this is a statistical fluctuation.

{
\def\bibname{\Large\bf References}
\def\refname{\Large\bf References}
\pagestyle{plain}
{\raggedright
\bibliography{./SR.bib}}

\providecommand{\etal}{et al.\xspace}
\providecommand{\coll}{Coll.\xspace}
\catcode`\@=11
\def\@bibitem#1{%
\ifmc@bstsupport
  \mc@iftail{#1}%
    {;\newline\ignorespaces}%
    {\ifmc@first\else.\fi\orig@bibitem{#1}}
  \mc@firstfalse
\else
  \mc@iftail{#1}%
    {\ignorespaces}%
    {\orig@bibitem{#1}}%
\fi}%
\catcode`\@=12
\begin{mcbibliography}{10}

\bibitem{epj:c12:35}
ZEUS Coll., J.~Breitweg \etal,
\newblock Eur.\ Phys.\ J.{} {\bf C~12},~35~(2000)\relax
\relax
\bibitem{zfp:c72:593}
H1 Coll., C.~Adloff \etal,
\newblock Z.\ Phys.{} {\bf C~72},~593~(1996)\relax
\relax
\bibitem{pl:b407:402}
ZEUS Coll., J.~Breitweg \etal,
\newblock Phys.\ Lett.{} {\bf B~407},~402~(1997)\relax
\relax
\bibitem{pl:b293:465}
ZEUS Coll., M.~Derrick \etal,
\newblock Phys.\ Lett.{} {\bf B~293},~465~(1992)\relax
\relax
\bibitem{zfp:c63:391}
ZEUS Coll., M.~Derrick \etal,
\newblock Z.\ Phys.{} {\bf C~63},~391~(1994)\relax
\relax
\bibitem{nim:a361:197}
U.~Bassler and G.~Bernardi,
\newblock Nucl.\ Inst.\ Meth.{} {\bf A~361},~197~(1995)\relax
\relax
\bibitem{misc:eps01:493}
ZEUS Coll.,
\newblock {\em {$D^{\ast \pm}$} Production in Deep Inelastic Scattering},
  2001\relax
\relax
\bibitem{proc:hera:1991:23}
S.~Bentvelsen, J.~Engelen and P.~Kooijman,
\newblock in {\em Proc.\ Workshop on Physics at {HERA}}, eds.~W.~Buchm\"uller
  and G.~Ingelman, Vol.~1, p.~23.
\newblock Hamburg, Germany, DESY, 1992\relax
\relax
\bibitem{cpc:67:465}
G.~Marchesini \etal,
\newblock Comp.\ Phys.\ Comm.{} {\bf 67},~465~(1992)\relax
\relax
\bibitem{cpc:86:147}
H.~Jung,
\newblock Comp.\ Phys.\ Comm.{} {\bf 86},~147~(1995)\relax
\relax
\bibitem{pl:b353:535}
B.W.~Harris, J.~Smith,
\newblock PL{} {\bf B~353},~535~(1995)\relax
\relax
\bibitem{pr:d57:2806}
B.W.~Harris, J.~Smith,
\newblock PR{} {\bf D~57},~2806~(1995)\relax
\relax
\bibitem{pr:d27:105}
C.~Peterson \etal,
\newblock Phys.\ Rev.{} {\bf D~27},~105~(1983)\relax
\relax
\bibitem{hep-ex-9912064}
L.~Gladilin,
\newblock Preprint \mbox{hep-ex/9912064}, 1999\relax
\relax
\bibitem{epj:c5:461}
M.~Gl\"uck, E.~Reya and A.~Vogt,
\newblock Eur.\ Phys.\ J.{} {\bf C~5},~461~(1998)\relax
\relax
\bibitem{epj:c12:375}
CTEQ Coll., H.L.~Lai \etal,
\newblock Eur.\ Phys.\ J.{} {\bf C~12},~375~(2000)\relax
\relax
\bibitem{epj:c7:609}
ZEUS Coll., J.~Breitweg \etal,
\newblock Eur.\ Phys.\ J.{} {\bf C~7},~609~(1999)\relax
\relax
\end{mcbibliography}
}

\vfill\eject

%\newpage

\begin{figure}
  \begin{center}
    \begin{minipage}[tl]{0.40\linewidth}
      \epsfig{file=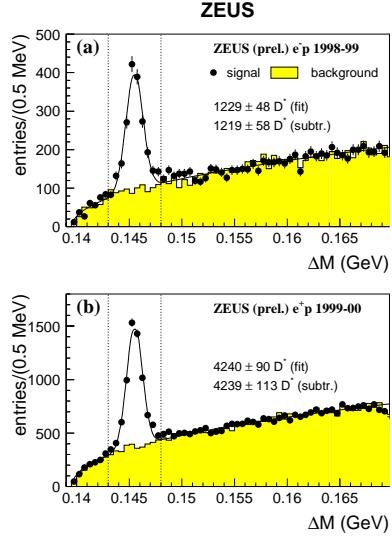, scale=0.2, width=\linewidth}
    \end{minipage}\hfill
    \begin{minipage}[tr]{0.45\linewidth}
      \caption{\textit{Data (solid dots) for $\Delta M = (M_{K \pi \pi_s} - M_{K \pi})$ for $e^-p$ data above and $e^+p$ data below.  The background from wrong charge combinations in shown as the filled histogram.  The solid line shows the result of the fit described in the text; the dashed vertical line indicates the signal region.}}
    \end{minipage}
  \end{center}
\end{figure}

\begin{figure}
  \begin{center}
    \begin{minipage}[b]{0.45\linewidth}
      \epsfig{file=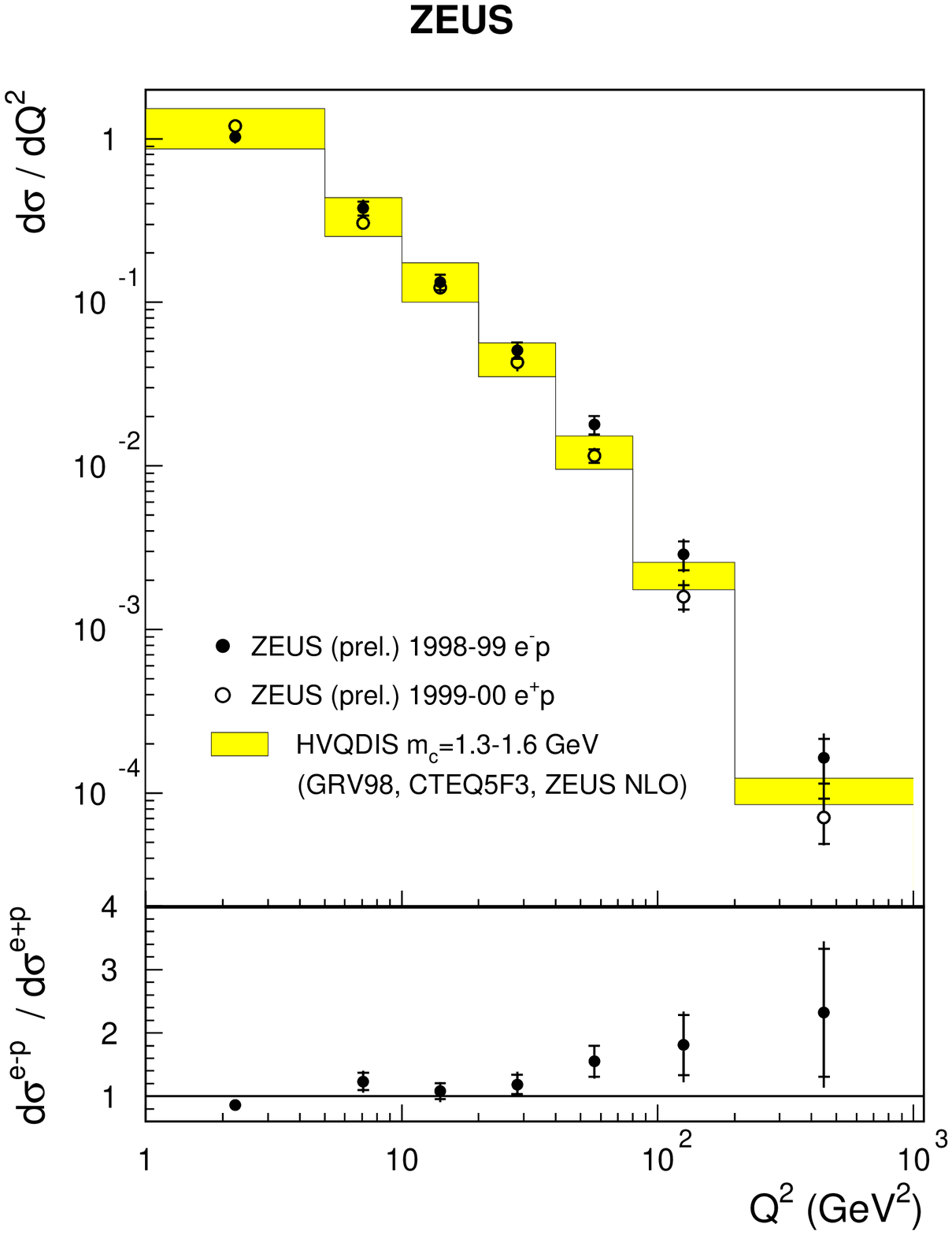, scale=0.35, width=\linewidth}
    \end{minipage}\hfill
    \begin{minipage}[b]{0.45\linewidth}
      \epsfig{file=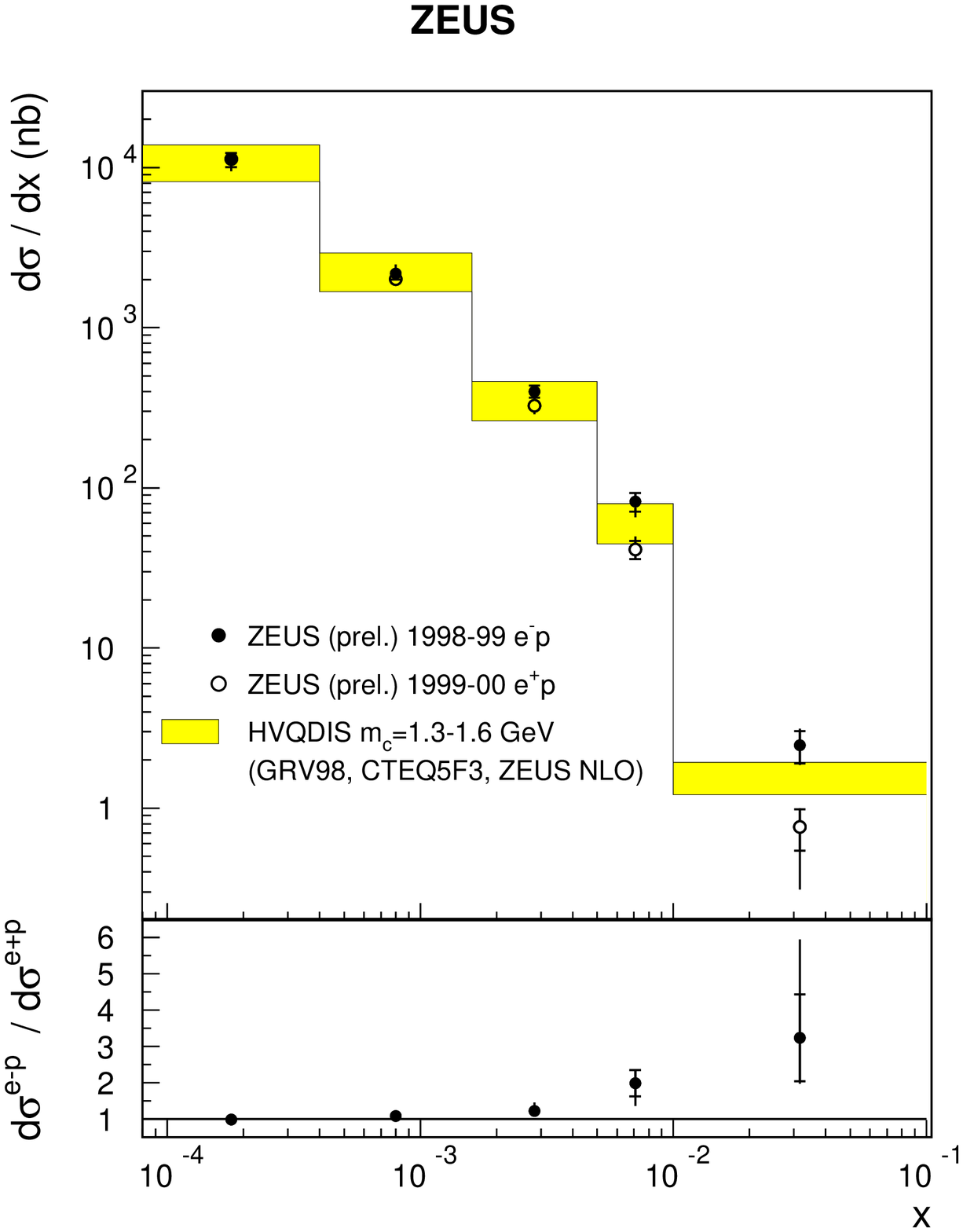, scale=0.35, width=\linewidth}
    \end{minipage} \caption{\textit{Differential $D^{\ast\pm}$ cross
        sections for $e^-p$ and $e^+p$ data as a function of $Q^2$ on
        the left and $x$ on the right, compared to the NLO QCD
        calculation of HVQDIS.  The inner error bars show the
        statistical uncertainties, while the outer ones are the
        statistical and systematic errors added in quadrature.  The
        boundaries of the shaded band for the HVQDIS prediction
        correspond to the full uncertainty due to the charm mass
        variation and choice of structure function as described in the
        text.  The lower portion of each plot shows the ratio of the
        $e^-p$ to $e^+p$ cross sections.}}  \end{center} \end{figure}
\end{document}